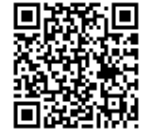

*Research Article*

# The Effect of Young People Not In Employment, Education or Training, On Poverty Rate in European Union

**Ionuț JIANU**

Bucharest University of Economic Studies, Bucharest, Romania

ionutjianu91@yahoo.com

## Abstract

This paper aims to estimate the effect of young people who are not in employment, education or training (neets rate) on the people at risk of poverty rate in the European Union. Statistical data covering the 2010-2016 period for all EU-28 Member States have been used. Regarding the methodology, the study was performed by using Panel Estimated Generalized Least Squares method, weighted by Period SUR option. The effect of neets rate on poverty rate proved to be positive and statistically significant in European Union, since this indicator includes two main areas which are extremely relevant for poverty dimension. Firstly, young unemployment rate was one of the main channels through which the financial crisis has affected the population income. Secondly, it accounts for the educational system coverage and its skills deficiencies.

**Keywords**: poverty, education, employment, social, panel

**JEL Classification:** I24, I30, I32, J21

## Introduction

Poverty is one of the greatest challenges of the last decade. The upward trend of the poverty rate in the European Union was a consequence of the economic and financial crisis. The situation is difficult as unemployment; one of the factors that have an important role to play in the dynamics of poverty, is still high, even if some countries experienced labour market improvements.

Latest European studies show that this theme has a particular importance since 1 out of 4 European citizens' faces challenges related to poverty, social exclusion and material deprivation. According to a survey of the European Commission, 8 out of 10 Europeans believe that the main challenges of the European Union are unemployment, social inequalities and migration. In this context, poverty has become one of the most discussed topics in the last few years and the review of the channels by which poverty drivers are influencing its level is essential for identifying the priorities that decision-makers should focus on.

The motivation for choosing the theme lies in its actuality and in the fact that recent studies have focused more on the separate analysis of the factors that are included in the young people neither in employment nor in education or training (NEETs) rate and didn't asses the aggregate impact of this indicator on poverty. This concept has been used starting from 2010 to provide a descriptive picture of the challenges faced by young people and to streamline youth-oriented policies in the EU.

The main objective of the paper is to assess the impact of the NEETs on the people at risk of poverty rate. In order to meet this purpose, other explanatory variables were used, such as government spending on social protection and in-work poverty rate (people over 18 age).

**Literature Review**

The literature studying this concept provides some evidence on poverty drivers, but research findings are sometimes questionable as a result of the qualitative issues their estimates are facing. Existing literature in this area does not focus on studying the aggregate effect of NEETs that includes both unemployment and early leavers from education and training. Most studies focused on the separate analysis of these explanatory factors and sometimes it has obtained results contrary to the economic theory.

The World Bank (2005) has framed the determinants of poverty into four pillars, as follows: (i) regional characteristics; (ii) community characteristics; (iii) characteristics of households; (iv) individual characteristics. Individual characteristics take into account factors related to age, education, status on the labour market, health and ethnicity. Atkinson (2013) has shown that the increase in the poverty rate is also caused by national institutions as well as by the policies adopted on the labour market. On the other hand, Duiella and Turrini (2014) have found that the impact of unemployment, long-term unemployment and GDP per capita on people at risk of poverty rate is not significant, it is the impact of long-term unemployment proving to be even negative.

Regarding the people at risk of poverty rate, some authors have stated that the cause of this type of poverty cannot be accurately identified, it could be due to: low hourly wages, too few working hours or recurrent periods of unemployment (Crettaz, 2011; Larsson and Halleröd, 2011). However, other researchers have shown that there is a low correlation between low salary levels and in-work poverty rate, most low wage earners are not exposed to the risk of poverty (Corluy and Vandenbroucke, 2014; Marx and Nolan, 2014).

Lohman (2009) and Crettaz (2011) have developed an analysis demonstrating that in-work poverty is not caused by the low level of wages and is a consequence of the high number of family members. Other studies have shown that people in Northern member states may be closer to the risk of poverty than those from other EU countries because they leave their parents' homes at a lower age and no longer benefit from their money support during the transition period from student to employee (Halleröd and Ekbrand, 2014).

Notten and Guio (2016) analysed the relationship between social transfers and material deprivation and identified a strong negative impact of this social tool. They also found that social transfers have a high capacity to reduce the number of people falling into the category of severe material deprivation. In particular, the literature in this field has identified a negative correlation between poverty rate and social spending (Behrendt, 2002). Kühner (2007) highlighted the limitations of this indicator, as it may react to changes in the unemployment rate as a result of cyclical factors.

**Methodology**

In this section, the methodology used is presented in order to estimate the impact of the young people who are neither in employment nor in education or training rate on the people at risk of poverty rate in the European Union (relative poverty). In order to obtain an aggregate impact, panel data with annual frequency is used.

Firstly, the indicators mentioned below and published by Eurostat for all EU member states, covering the period of 2010-2016 (due to the limited availability of time series for some European countries) was extracted.

All necessary operations for the estimation of the impact of the NEETs rate on poverty rate were conducted using Eviews 9.0 software. Further, the stationarity for the panel data was checked using "Summary" window which provides a detailed view of the results of the following stationarity tests:

✓  Assuming common unit root process (null hypothesis: unit root / alternative: no unit root):

  • Levin, Lin & Chu t* (applied in the following assumptions: trend and constant, constant, absence of trend and constant) - 3 results - some disadvantages of the test are: (a) if the number of observations per cross-section is small, the power of the test may be questionable; (b) this test ignores the possibility of the cross-section dependence.
  • Breitung t-stat - (applied if the test equation includes the trend and constant) - 1 result - this test differs from Levin, Lin & Chu t* since only the autoregressive portion (and not the exogenous components) is removed when constructing the standardized proxies.

✓  Assuming individual unit root process (null hypothesis: unit root / alternative: some cross-sections without unit root):

  • Im, Pesaran and Shin W-stat (applied in the following assumptions: trend and constant, constant) - 2 results - this test works better with low number of observations per cross-section than Breitung and has little power when trend is included in the analysis;
  • ADF - Fisher Chi-square (applied in the following assumptions: trend and constant, constant, absence of trend and constant) - 3 results - this test allows each cross-section to have a different lag length;
  • PP - Fisher Chi-square (applied in the following assumptions: trend and constant, constant, absence of trend and constant) - 3 results.

The stationarity hypothesis was confirmed when more than half of the total results (12) indicated this. The approach followed was suitable for this analysis since it provides a broader view on the stationarity process, while the use of a single test assuming common unit root process may return inappropriate results as panel data could be exposed to the heterogeneity risk. On the other hand , the homogeneity in panel data may facilitate the persistence of the correlation between cross-sections (eg. the same impact of the autoregressive term on the endogenous) which rejects the assumption of individual unit root process. In order to select the optimal lag, "Schwarz information criterion (SIC)" was used which was calculated by the following formula:

*Schwarz information criterion* $= n ln(n - p - 1)^{-1} \sum_{i=1}^{n} \varepsilon_t^2 + n^{-1} p ln(n)$                    (1)

---

_______________________________________________________________________

, where n is the sample size, ln is natural logarithm and $\varepsilon_t$ are the residuals. When using a maximum probability estimate for parameter estimation, there may be a risk of over-fitting as a consequence of the increase in additional parameters. The Schwarz criterion restricts stronger the additional parameters than the Akaike criterion, both are the most used criteria for lag selection in the relevant economic literature. In addition, if the number of observations per cross-section is greater than the number of exogenous variables, the criteria estimates are consistent and impartial. Following several estimates with different lags, the lag associated with the lowest SIC value was chosen.

The variables used proved to be stationary at level, which indicates using the EGLS - Estimated Generalized Least Squares method. The problem of heteroskedasticity of the residuals, autocorrelation and the existence of general correlations between the cross sections, has required the application of the "Period SUR" option on the following equation:

$$povertyrate = \alpha_0 + \beta_0 inworkpovertyrate + \beta_1 NEETsrate + \beta_2 socialexp + \varepsilon_t \qquad (2)$$

, where:

➤ *povertyrate* - people at risk of poverty rate, after social transfers (% - the share of the population earning less than 60% of the median equivalised national income after social transfers);
➤ *inworkpovertyrate* - the percentage of the employment at risk of poverty, after social transfers (% - the share of the employees earning less than 60% of the median equivalised national income);
➤ *NEETsrate* - the percentage of young people neither in employment nor in education or training systems (% - this category includes young people aged 15-24 that are outside employment, education systems or have not participated in training programs in the last 4 weeks preceding the survey);
➤ *socialexp* - social government expenditures (expressed as a percentage of GDP).

A number of 196 observations resulted from the time series used for the 28 EU member states analysed. Also, the following econometric tools have been used in order to validate the maximum verisimilitude of the estimators (*Table 1*) :

**Table 1: Econometric tools used for validating maximum verisimilitude of the estimators**

| Tool used | Hypothesis checked |
|---|---|
| Fisher test | Valid / invalid model |
| Jarque-Bera test | Normally / abnormally distribuited residuals |
| Durbin-Watson test | Absence / existence of autocorrelation |
| Breusch-Pagan-Godfrey test | Heteroskedasticity / Homoskedasticity |
| Cross-section Dependence Test (Breusch-Pagan, Pesaran CD, Pesaran scaled LM) | Absence / existence of cross-section dependence |
| Pearson correlation | Existence / absence of multicolinearity |

*Source: Own processings using Microsoft Office Word 2016*

In order to test the heteroskedasticity / homoskedasticity, the Breusch-Pagan-Godfrey test was applied. First, the following equation was estimated:

___________

_______________________________________________________________

$$residual^2 = \lambda_0 + \mu_0 inworkpovertyrate + \mu_1 NEETsrate + \mu_2 socialexp + \varepsilon_t \qquad (3)$$

, where $residual^2$ represents the square of the residuals of equation (2).

Further, the probability of the Breusch-Pagan-Godfrey test was estimated by applying the CHISQ.DIST.RT function which provides the one-tailed probability of the chi-squared distribution for the following arguments: (i) the product of the number of observations and the R-squared value associated with the equation (3) and (ii) the number of exogenous variables, excluding the constant (degrees of freedom).

**Results and Interpretations**

In this section, the main results of the empirical analysis carried out have been displayed including the developments of the variables and the results of the estimation presented in the methodology. The rate of people at risk of poverty (earning less than 60% of the median equivalised national income), increased by less than 1 percentage point in 2010-2016 period (0.8 percentage points) - *Figure 1*. However, the evolution of the poverty rate in the EU member states was extremely divergent, with some of them reporting increases by more than 2 percentage points (Luxembourg, Hungary, Bulgaria, Netherlands, Romania, Estonia), and other countries recording falls by more than 1 percentage point (Croatia, United Kingdom, Denmark, Finland).

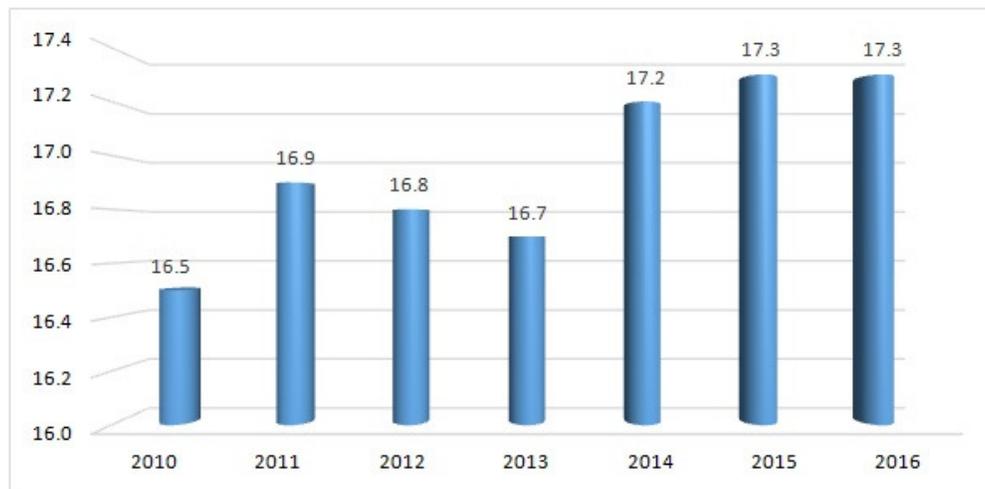

**Figure 1: Evolution of the people at risk of poverty rate in EU**

*Source: Own processings using Eurostat database*

The highest increases in the poverty rate in 2016 compared to 2010 were recorded in the Netherlands (2.4 pp), Romania (3.7 pp) and Estonia (5.9 pp). At the opposite end were United Kingdom (-1.2 pp), Denmark (-1.4 pp) and Finland (-1.5 pp). The unfavorable developments of this indicator can be attributed to the government spending on social protection (often inefficient), lack of structural reforms, low labour productivity in line with low wage earnings, poor quality of tertiary education

_______________

systems and unsustainable economic growth. A significant impact on it had also been exercised by the economic and financial crises that negatively influenced the population income, mainly as a result of its interaction with the increasing trend of the unemployment rate. Moreover, unemployment hit hardest the categories of people with low incomes, as people earning high wages were able to orient their financial resources to higher yielding economic activities. The impact was lower in the resilient economies that have used appropriate tools for shock absorption.

In 2016, the highest poverty rates were recorded in Romania (25.3%), Bulgaria (22.9%) and Spain (22.3%), while the Czech Republic (9.7%) , Finland (11.6%) and Denmark (11.9%) recorded the lowest levels of this indicator. Among them, Romania (26th place) and Bulgaria (27th place) occupy the last positions in the EU in a ranking made by the Development Finance International Group and the Oxfam International Confederation regarding the commitment to reduce inequality. They also occupy the last two positions in the EU regarding the commitment of national governments to make the neccesary health, education and social protection expenditures. Although the concept of inequality and poverty are different, these have some common bases, such as the use of similar social policy instruments or their linking with economic growth. Evolution of the statistical data covering 2010-2016 period shows a positive correlation of 87.39% between the evolution of the Gini coefficient and the people at risk of poverty rate. Empirical evidences expressed the necessity of the assessment of the relationship between the poverty rate and the following indicators: in-work poverty, NEETs rate (which includes both early leavers from education systems and unemployment) as well as the social protection expenditures of general government (expressed as a percentage of GDP).

According to *Figure 2*, there is a strong link between the poverty rate after social transfers and the in-work poverty rate. This conclusion is predictable given that in-work poverty is also a component of the indicator under review. This can also be observed by studying the Panel - Pearson - high correlation coefficient (74.23%) or the R-squared value (55.35%).

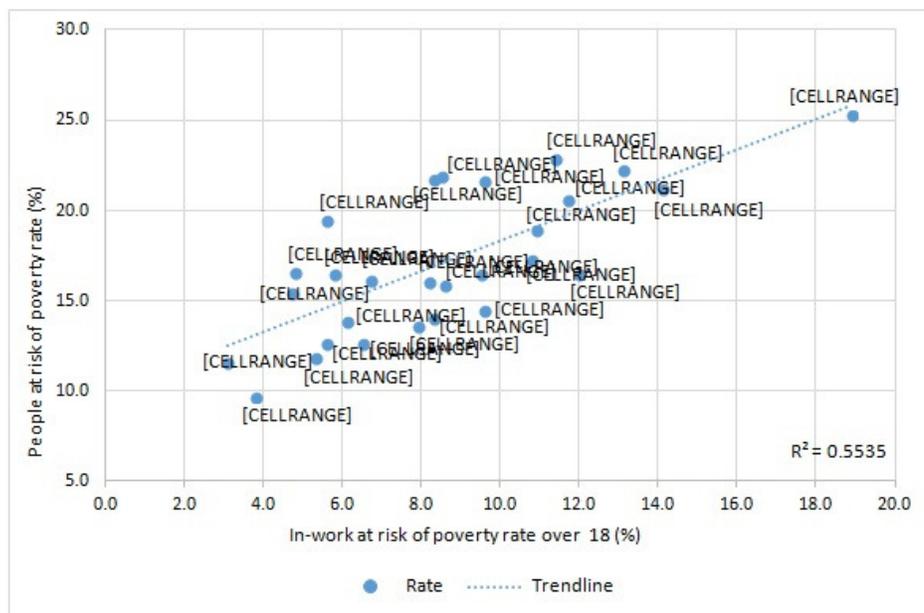

**Figure 2: The relationship between people at risk of poverty rate and in-work at risk of poverty rate over 18 in EU (2016)**

*Source: Own processings using Eurostat database*

However, the results of such a method could be subject to statistical uncertainty, which is why a more compact form of the model was designed. In-work poverty rate was included in the model to increase the accuracy of the model given its control variable character.

As it can be seen, Romania recorded in 2016 both the highest people at risk of poverty rate after social transfers and the highest rate of the employed population over 18 years at risk of poverty, which shows that the main cause of poverty poverty in Romania consists in the low level of wages. Largely, there is a similarity between the positions of these indicators in the EU. Finland, the Czech Republic, Denmark recorded the lowest rates of the two indicators mentioned, while Romania and Spain recorded the highest rates of them. One of the countries that recorded a high in-work poverty rate (the 4th rate in the EU - 12.0%) and a low people at risk of poverty rate after social transfers (13th rate of the EU - 16.5%, below the EU average of 17.3%) is Luxembourg. The reason for this inconsistency lies in the fact that this country is the most important Europe's financial centre; a significant share of population income is obtained by participating to the economic activities of financial market.

Also, the in-work poverty rate increased by 1.3 percentage points in 2016 compared to its 2010 level in EU, higher than the one recorded by the poverty rate after social transfers, highlighting the fact that the population of the member states starts to orient their savings towards the capital market in order to obtain additional gains that have a higher return.

*Figure 3* highlights a negative correlation (Panel Pearson correlation = -35.14%) between the evolution of government spending on social protection (% of GDP) and that of the people at risk of poverty rate, which makes feasible a negative impact of the government expenditures on social protection on poverty rate. Also, the level of government spending on social protection in the EU explains approximately only a sixth of the poverty rate development.

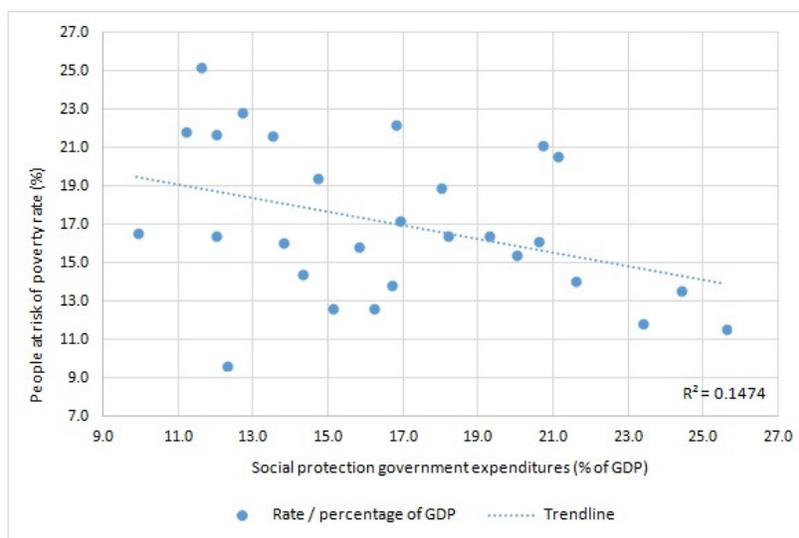

**Figure 3: The relationship between people at risk of poverty rate and social protection government expenditures in EU (2016)**

*Source: Own processings using Eurostat database*

After rising to 19.5% of GDP in 2013, government spending on social protection declined to 19.1% of GDP in 2016. This evolution was due to the budgetary constraints of some member states such as Ireland (-7.7 pp of GDP compared to the level recorded in 2010), Hungary (-3.1 pp of GDP), Lithuania (-2.9 pp of GDP). On the other hand, Finland has increased its social spending share in GDP by 2.8 percentage points, followed by Greece (1.9 pp of GDP - which continues to be a major supporter of social policy despite their significant challenges they are facing which are related to the high public debt) and Cyprus (1.6 pp of GDP). Romania, although in 2010 recorded the fifth smallest share in GDP from the EU in this type of spending, chose to reduce the government spending on social protection during this period by 2.3 pp of GDP, this is the fourth highest cut in the EU. In 2016, Romania recorded the third smallest share in GDP of these expenditures (11.6% of GDP), which was only higher than the one reported by Lithuanian (11.2% of GDP) and Irish (9.9% of GDP) authorities. In Ireland, public debt fell sharply from 119.6% of GDP in 2012 to 68% of GDP in 2017. However, given its limited fiscal options due to the high public debt, the Irish authorities have proposed reaching a 45% of GDP government debt by 2025. The budgetary situation is one of the main causes of the downward trend in government spending on social protection in Ireland.

Regarding the relationship between poverty rate and NEETs rate, *Figure 4* shows a high correlation between them, the coefficient of Pearson correlation reaching 63.61%. The NEETs rate is a more complex indicator that takes into account both labour market factors and aspects related to the education system and student motivation. Consequently, this indicator explains about 37% of the developments in people at risk of poverty rate in EU.

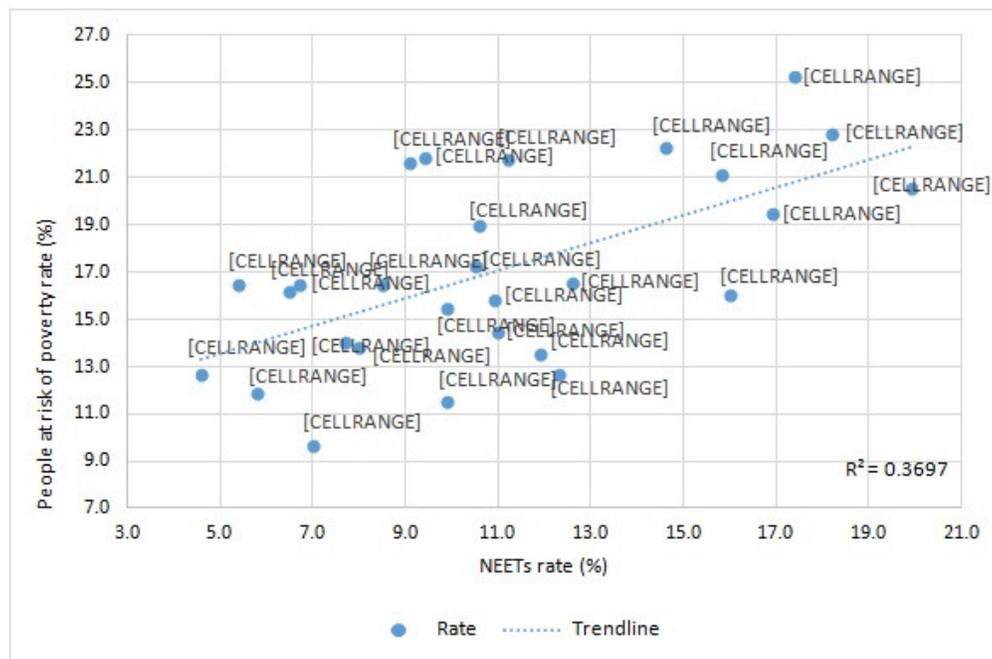

**Figure 4: The relationship between people at risk of poverty rate and NEETs rate in EU (2016)**

*Source: Own processings using Eurostat database*

The impact of the economic and financial crises shock from 2009 on young Europeans' participation on labour market or in education systems was quite strong. From the value of NEETs rate of 10.9% in 2008, it peaked in 2012 to 13.2%, being followed by a downward trend until reaching the level of 11.6% in 2016.

According to the developments from 2010-2016 period, the largest cuts in NEETs rate occurred in Estonia (-4.9 pp), Latvia (-6.6 pp) and Ireland (-6.8 pp). The crisis had a higher impact on the status of young people on labour market and education system in countries like Cyprus (4.3 pp), Croatia (1.2 pp) and Greece (1.0 pp). In 2016, the highest NEETs rates were recorded in Romania (17.4%), Bulgaria (18.2%) and Italy (19.9%), while in Denmark (5.8%) , Luxembourg (5.4%) and the Netherlands (4.6%) reported the lowest levels of this indicator.

Eurofund carried out an analysis through which the causes of NEETs and its structure in the EU are reviewed. Therefore, Eurofound identified the following NEETs categories:

- ✓ 7.8% - young re-entrants on labour market or education systems who will no longer be taken into account by this indicator;
- ✓ 29.8% - young people facing short-term unemployment;
- ✓ 22% - young people facing long-term unemployment;
- ✓ 6.8% - young people with disabilities;
- ✓ 15.4% - young people with family responsibilities (eg childcare);
- ✓ 5.8% - young discouraged people;
- ✓ 12.5% - other young persons.

As it can be seen, the NEETs rate includes a significant number of socially vulnerable people. The analysis of this inidcator at granular level provides a clearest view of the positive relationship between NEETs rate and people at risk of poverty.

Further, the impact of NEETs rate on people at risk of poverty rate is estimated. Following the estimation of the model (*Figure 5*), it was found that the estimators are significant, which creates the premises for a high degree of confidence in the resulting coefficients, the probabilities associated with them are all below 5%. Also, the hypothesis of the Gauss-Markov theorem stating that the standard errors must be non-but-close to zero in order to confirm the maximum estimator's verisimilitude was confirmed.

Coefficients were interpreted in line with the "caeteris-paribus" hypothesis. According to the results, the increase by 1 pp of in-work (over 18 years) at risk of poverty rate leads to an increase in the poverty rate after social transfers by 0.559 pp. This relationship derives from the fact that this indicator is a component of the endogenous variable, as was discussed above.

Regarding the impact of government spending on social protection, raising it by 1 percentage point of GDP leads to a decline in the poverty rate by 0.181 pp. A major cause would be the function of these expenditures to cover the material deprivation of the population in order to facilitate a decent living standard for actual and further generations. This type of expenditure gives the possibility for low earners to overcome the poverty line, set at 60% of the median equivalised national income.

Returning to the main objective of the paper, it was found that the 1 pp increase in the NEETs rate leads to an increase in the rate of people at risk of poverty after social transfers by 0.135 pp. This effect is caused by the income pressure challenges generated by unemployment or school drop-out. The NEETs coefficient is lower than the other coefficients in absolute form given that the poverty rate takes into account all age groups, while NEETs rate relies on the 15-24 age group.

In order to accept the maximum verisimilitude of the estimators, it was necessary to check the hypotheses of the Gauss-Markov theorem. According to the probality of the Fisher test, the model is statistically valid and the coefficient of determination indicates that 50.74% of the poverty rate fluctuation comes from the dynamic of the exogenous variables.

Next, the residuals testing the procedure were started by checking the autocorrelation of the residuals. The result of the Durbin-Watson test (1.951906) range between DU (1.79688) and 4-DU (2.20312) statistics, which confirmed the absence of autocorrelation, mentioning that DL (1.73445) and DU statistics for a total of 196 observations and 4 explanatory variables were used (including the constant) at a significance degree of 5%.

| Dependent Variable: POVERTYRATE | | | | |
|---|---|---|---|---|
| Method: Panel EGLS (Period SUR) | | | | |
| Date: 08/05/18   Time: 20:01 | | | | |
| Sample: 2010 2016 | | | | |
| Periods included: 7 | | | | |
| Cross-sections included: 28 | | | | |
| Total panel (balanced) observations: 196 | | | | |
| Linear estimation after one-step weighting matrix | | | | |
| Period SUR (PCSE) standard errors & covariance (d.f. corrected) | | | | |
| Variable | Coefficient | Std. Error | t-Statistic | Prob. |
| INWORKPOVERTYRATE | 0.559232 | 0.047505 | 11.77196 | 0.0000 |
| SOCIALEXP | -0.181560 | 0.063323 | -2.867219 | 0.0046 |
| NEETSRATE | 0.135345 | 0.039473 | 3.428753 | 0.0007 |
| C | 13.35735 | 1.241003 | 10.76335 | 0.0000 |
| | Weighted Statistics | | | |
| R-squared | 0.507414 | Mean dependent var | | 2.405714 |
| Adjusted R-squared | 0.499718 | S.D. dependent var | | 3.528554 |
| S.E. of regression | 0.950488 | Sum squared resid | | 173.4582 |
| F-statistic | 65.92667 | Durbin-Watson stat | | 1.951906 |
| Prob(F-statistic) | 0.000000 | | | |
| | Unweighted Statistics | | | |
| R-squared | 0.664053 | Mean dependent var | | 16.57704 |
| Sum squared resid | 874.8612 | Durbin-Watson stat | | 0.108514 |

**Figure 5: Estimation results**
*Source: Own processings using Eviews 9.0*

Figure 6 highlights the result of the Jarque-Bera test and its probability, which is higher than 5% (29.95%) and confirms the null hypothesis of normally distributed residuals.

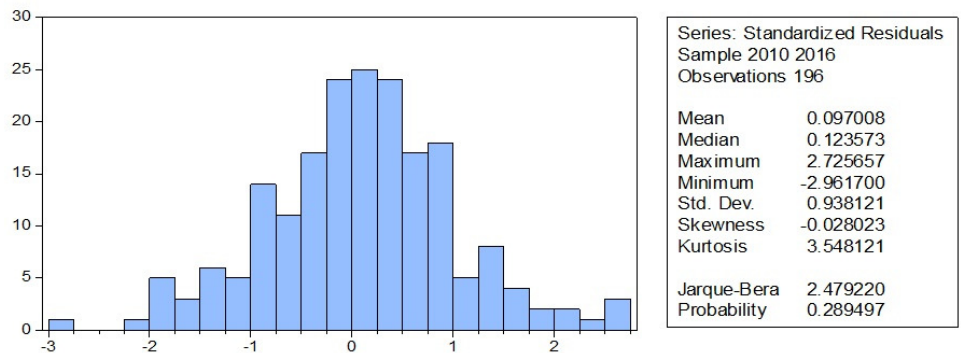

**Figure 6: Histogram - Normality test**

*Source: Own processings using Eviews 9.0*

According to *Figure 7*, all results associated with the performed tests (Breusch-Pagan LM, Pesaran LM, Pesaran CD) returned the probabilities of over 5%, which led to the acceptance of the null hypothesis according to which there is no dependence between cross-sections. *Figure 8* highlights the probability of the Breusch-Pagan-Godfrey test of 5.49%, which confirmed the homoskedastic feature of the model, as it exceeds the significance degree of 5%.

| Residual Cross-Section Dependence Test | | | |
|---|---|---|---|
| Null hypothesis: No cross-section dependence (correlation) in weighted residuals | | | |
| Equation: EQPOVERTY01 | | | |
| Periods included: 7 | | | |
| Cross-sections included: 28 | | | |
| Total panel observations: 196 | | | |
| Note: non-zero cross-section means detected in data | | | |
| Cross-section means were removed during computation of correlations | | | |
| | | | |
| Test | Statistic | d.f. | Prob. |
| | | | |
| Breusch-Pagan LM | 400.3456 | 378 | 0.2057 |
| Pesaran scaled LM | -0.205647 | | 0.8371 |
| Pesaran CD | -0.375082 | | 0.7076 |
| | | | |

**Figure 7: *Cross-section dependence test***

*Source: Own processings using Eviews 9.0*

| Dependent Variable: RESIDUAL^2 | | | | |
|---|---|---|---|---|
| Method: Panel Least Squares | | | | |
| Date: 08/05/18  Time: 20:19 | | | | |
| Sample: 2010 2016 | | | | |
| Periods included: 7 | | | | |
| Cross-sections included: 28 | | | | |
| Total panel (balanced) observations: 196 | | | | |
| | | | | |
| Variable | Coefficient | Std. Error | t-Statistic | Prob. |
| | | | | |
| INWORKPOVERTYRATE | -0.023548 | 0.032988 | -0.713836 | 0.4762 |
| SOCIALEXP | -0.065685 | 0.026859 | -2.445590 | 0.0154 |
| NEETSRATE | 0.018240 | 0.024084 | 0.757345 | 0.4498 |
| C | 1.975632 | 0.631818 | 3.126899 | 0.0020 |
| | | | | |
| R-squared | 0.038797 | Mean dependent var | | 0.884991 |
| Adjusted R-squared | 0.023778 | S.D. dependent var | | 1.409853 |
| S.E. of regression | 1.392990 | Akaike info criterion | | 3.520979 |
| Sum squared resid | 372.5610 | Schwarz criterion | | 3.587880 |
| Log likelihood | -341.0560 | Hannan-Quinn criter. | | 3.548064 |
| F-statistic | 2.583198 | Durbin-Watson stat | | 1.965747 |
| Prob(F-statistic) | 0.054628 | | | |
| | | | | |

| Heteroskedasticity test | |
|---|---|
| R-squared | 0.038797 |
| Number of observations | 196 |
| Degrees of freedom | 3 |
| Breusch-Pagan-Godfrey probability | 0.054943 |

**Figure 8: Heteroskedasticity test**

*Source: Own processings using Eviews 9.0 and Microsoft Office Excel 2016*

Finally, *Table 2* shows the low correlation between the exogenous variables, the maximum correlation is 38.06% and established between NEETs rate and in-work poverty. However, Klein's criterion was considered respected and the absence of multicollinearity was accepted.

Ultimately, the results obtained led to the validation of the accuracy of the estimators.

**Table 2: *Independent variables correlation matrix***

| Correlation matrix | NEETs rate | inworkpovertyrate | socialexp |
|---|---|---|---|
| NEETs rate | 1.00000 | 0.38057 | -0.26423 |
| inworkpovertyrate | 0.38057 | 1.00000 | -0.16576 |
| socialexp | -0.26423 | -0.16576 | 1.00000 |

*Source: Own processings using Eviews 9.0*

## Conclusions

The analysis confirms the existence of a positive relationship between the NEETs rate and the people at risk of poverty rate. The impact of the NEETs on poverty has been smaller in absolute terms than the impacts of the other factors analyzed, but the situation of young people in this category should be on the list of key policy priorities to reduce the poverty rate, given that the impact of certain government spending on poverty is quite different between member states and depends on psycho-cultural factors too.

On the other hand, a high positive impact of in-work poverty rate (over 18 age) on people at risk of poverty have been demonstrated. Although the relationship is quite intuitive, some developments at EU level are surprising, such as the higher increase in-work poverty rate than the one of people at risk of poverty rate after social transfers, which highlights new labour market imbalances related to low wage earnings obtained by some social classes, or by the high number of members in households. For countries recording high in-work poverty rates, the implementation of structural reforms that contribute to the development of the human factor is essential, including improving the quality of the education and health system and the development of family policies. Moreover, a higher attention should be paid to the minimum wage setting policies. In this context, a differentiated minimum wage depending on the specificities of the economic sectors could be useful, but such a reform can be difficult to implement because regional administrations should receive more power in the decision-making process.

Stimulating the creation of new jobs will not solve the issue of NEETs and, therefore, of poverty. It is necessary to improve the quality of education and training systems as well as to improve social protection systems without discouraging the labour market participation. In this context, developing long-term reforms in order to guide the human mentality towards a participatory one (using instruments such as household minimum income) is essential.

___________